\documentclass[amsmath,amssymb,reprint,bibnotes]{revtex4-2}
\usepackage{graphicx}
\usepackage{bm}
\usepackage{amsmath} 
\usepackage{graphics}
\usepackage[dvipsnames]{xcolor}
\usepackage[mathlines]{lineno}
\usepackage{braket}
\usepackage{subfigure}
\usepackage[compact]{titlesec}
\begin{document}
\title{Active tuning of ENZ resonances in meta-antenna through phase modulation of optical pulse}
\author{Elif Ozturk$^{\bf (1,2)}$}
\author{Hira Asif$^{\bf (3,4)}$}
\author{Mehmet Gunay$^{\bf (5)}$}
\author{Mehmet Emre Tasgin$^{\bf (1)}$}
\author{Ramazan Sahin$^{\bf (3,4)}$}\email{rsahin@itu.edu.tr}

\affiliation{${\bf (1)}$ {Institute of Nuclear Sciences, Hacettepe University, 06800 Ankara, Turkey}}
\affiliation{${\bf (2)}$ {Department of Nanotechnology and Nanomedicine, Graduate School of Science and Engineering, Hacettepe University, 06800 Ankara, Turkey}}
\affiliation{${\bf (3)}$ {Department of Physics, Akdeniz University, 07058 Antalya, Turkey}}
\affiliation{${\bf (4)}$ {Türkiye National Observatories, TUG, 07058 Antalya, Turkey}}
\affiliation{${\bf (5)}$ {Department of Nanoscience and Nanotechnology, Faculty of Arts and Science, Burdur Mehmet Akif Ersoy University, Burdur 15030, Turkey}}

\date{\today}

\begin{abstract}
Plasmonic nanoantennas offer new avenues to manipulate the propagation of light in materials due to their near field enhancement and ultrafast response time. Here we investigate the epsilon-near-zero (ENZ) response in an L-shaped nanoantenna structure under the phenomenon of plasmonic analog of enhancement in the index of refraction. Using a quantum mechanical approach, we analyze the modulation in the response of probe field and emergence of ENZ frequency region both in the linear and nonlinear plasmonic system. We also demonstrate the active tuning of ENZ frequency region in a nanoantenna structure by modulating the phase of control pulse. The analytical and 3D FDTD simulation results show a significant spectral shift in the ENZ modes. Our proposed method offers the possibility to design and control optical tunable ENZ response in plasmonic metasurfaces without the use of ENZ material. Such metasurfaces can be used in on-chip photonic integrated circuits, further localization of incident fields, slow light operations and various quantum technologies.    
\end{abstract}

\maketitle
 
\section{Introduction}
The field of nanophotonics has witnessed remarkable advancements in recent years, driven by the possibility to manipulate light at subwavelength scales using nano-structured materials. Among these, metamaterials and nanoantennas have emerged as powerful platforms for controlling electromagnetic waves with unprecedented precision. A particularly intriguing phenomenon in this domain is the epsilon-near-zero (ENZ) effect, where the real part of permittivity of a material approaches zero [Re$\{\epsilon\}<<1$], enabling unique optical properties such as enhanced light-matter interactions, wavefront shaping, slow light propagation and extreme nonlinearities \cite{Deng2020, Kinsey2019}. These characteristics, combined with high tuning flexibility and CMOS compatibility \cite{Niu2018}, make ENZ materials highly competitive for applications in ultrafast optics, sensing, and energy harvesting \cite{reshef2019}.

The extraordinary response of ENZ material only occur at the small proximity of ENZ wavelength, refer as ENZ region. For optimal performance of the photonic systems, the ENZ wavelength needs to be align with optical source. Therefore, active tuning of ENZ resonances has become crucial to unlock real-time control over optical properties, paving the way for reconfigurable nanophotonic devices, switches, modulators, and even quantum networks \cite{Vasudev2013,Xie2020,Wood2018}.

Recent efforts have explored various tuning mechanisms on layered ENZ materials, including electrical gating, thermal modulation, and mechanical strain \cite{Qiu2021}.
\begin{figure}
\includegraphics[scale=0.5]{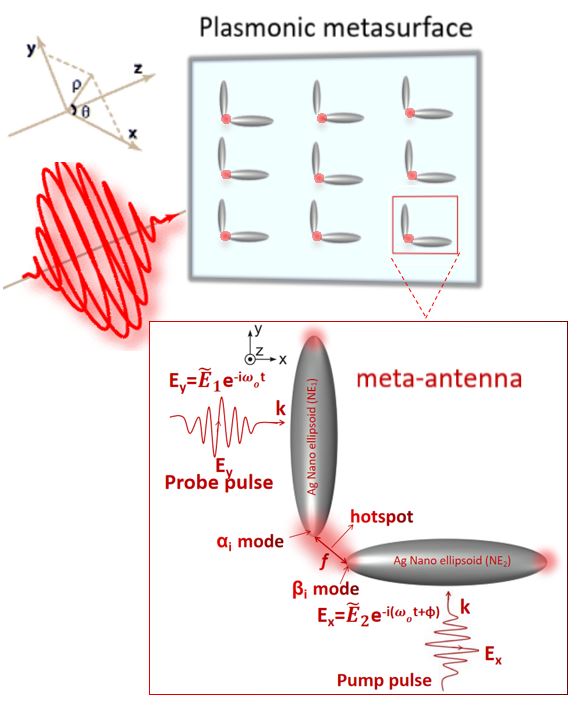}
\caption{\label{fig:1} Schematic diagram of plasmonic metasurface with nanoantenna structure supporting the ENZ response. In a pump-probe setup, an L-shaped plasmonic antenna}
\end{figure}  
For instance, electric tuning of indium tin oxide (ITO), a common ENZ material, demonstrate the electro-optic modulation with 70 nm optical bandwidth in infrared regime \cite{Gao2018, Kuang2019}. On the other hand, optical methods also offer temporary change in permittivity using time refraction phenomena \cite{HONG2020}. Nevertheless, these methods often suffer from slow response times, irreversibility, limited tunability, or complex fabrication requirements. Moreover, zero-index materials (ITO) and metals exhibit ENZ response in the near-infrared and ultraviolet region \cite{Campione2015}, which limits their applications in photonic systems operating in visible regime. Recently some studies have leveraged the unique properties of natural ENZ materials to tune plasmon resonances in ENZ-coupled plasmonic metasurfaces in the infrared regime \cite{Bilgi2020, manukyan2021,Kim2016}. While these studies have demonstrated the potential of active control of ENZ materials for tuning plasmon resonances, but tuning ENZ modes within plasmonic nanostructure is another challenge for realizing their full potential in ultrafast and reconfigurable photonic systems.

To address these challenges, here, we demonstrate a novel approach for active tuning of ENZ frequency in meta-nanoantenna through phase modulation of control pulse. By exploiting the interplay between the phase characteristics of incident light and the resonant modes of the nanoantenna, we demonstrate a dynamic and reversible tuning mechanism that operates at ultrafast timescales. In addition, such a system does not require natural ENZ material to couple with plasmonic metasurfaces giving it more liberty to employ in reversible photonic systems. 

We, here, utilized \textit{enhancement in the index of refraction} (EIR) phenomenon and explore how the phase-dependent characteristics of the pump influence the local field enhancement, tuning of spectral response, and realization of ENZ behavior in the meta-antenna structure for the first time to the best of our knowledge.
\begin{figure}[t!]
      \centering
        \includegraphics[height=2.8in]{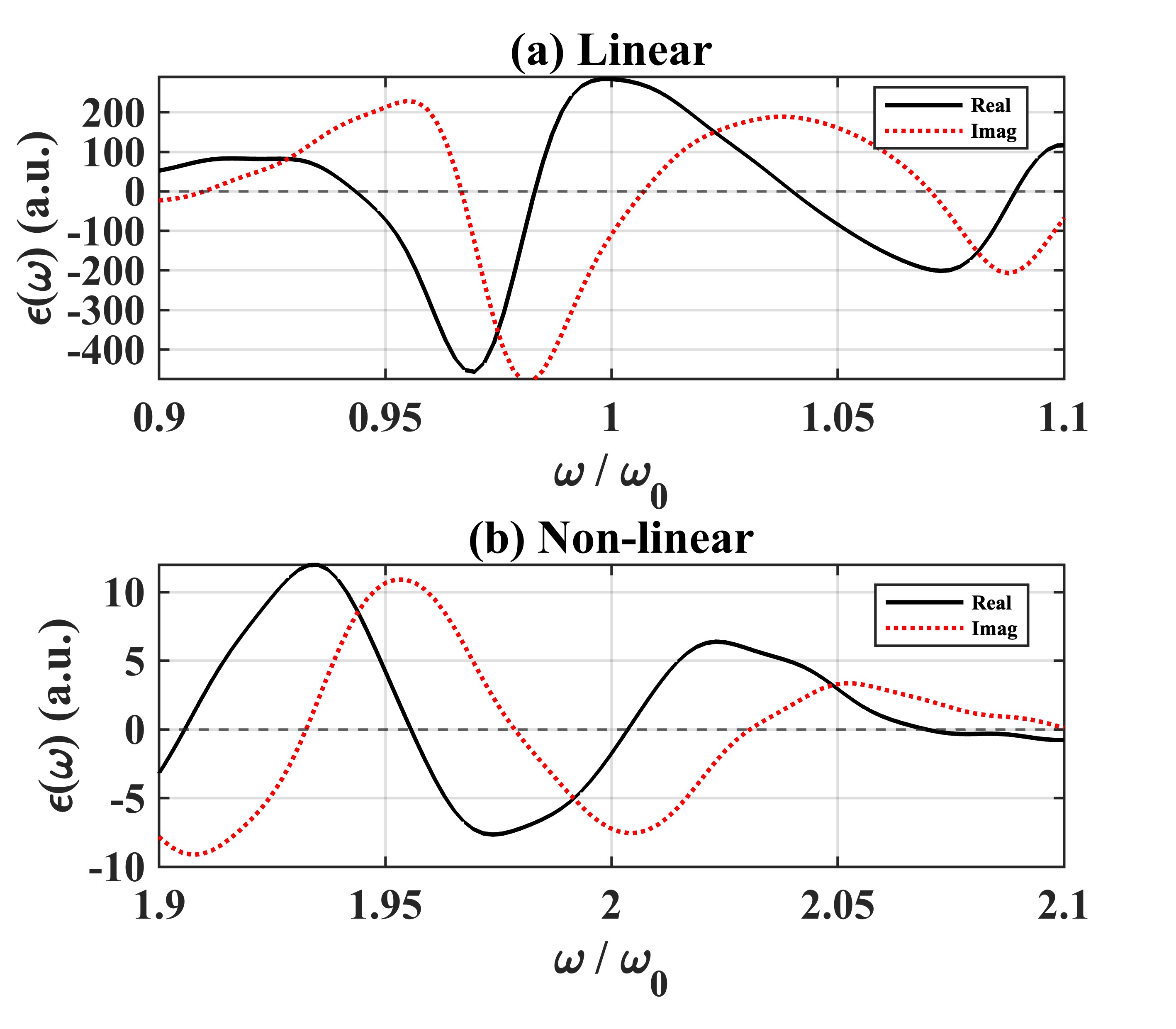}
    \caption{\label{fig:2} (a,b) Real and imaginary part of electric susceptibility in NE$_2$ demonstrating the enhanced EIR as a function of normalized frequency. At point A, absorption is zero and at A$^{,}$ the susceptibility [Re$\{\chi\}<-17$] becomes negative indicating the point of ENZ frequency.}
\end{figure}

Our model system compose of a plasmonic metasurface with periodic arrays of meta-antennas, excited by a circularly polarized pulse, as shown in Fig.\ref{fig:1}. Each nanoantenna is assembled in the form of L-shaped silver nano-ellipsoids (NEs), as previously studied in \cite{Panahpour2019, Yuce2021, Gunay2020}. In the pump-probe scheme, probe pulse E$_y$ excites linear plasmon mode ($\alpha_1$) in vertically aligned nanoantenna (NE$_1$), while pump pulse polarized in x-direction E$_x$ induces $\beta_1$ plasmon mode in horizontal ellipsoid NE$_2$. The frequency of pump and probe pulse is $\omega_o=3.19\times 10^{15}$ rad/s. The optical control in nanoantenna is governed by tuning phase of the pump pulse. To investigate the impact of pulse phase on the response function of probe field in both linear and nonlinear, we use a quantum mechanical approach and evaluate the change in EIR experienced by ($\alpha_i$) modes in vertical ellipsoid. Through theoretical modeling and finite difference time domain (FDTD) method, we demonstrate the feasibility of achieving precise and reversible tuning of ENZ resonance over a broad spectral range.\\ In the analytical approach, we derive the linear susceptibility experienced by the probe pulse ($\tilde{E}_1$) from the Hamiltonian \cite{Gunay2020} (see supplementary section (SS) for details), as follows, 
\begin{eqnarray}
\chi(\omega)= \frac{-if e^{-i\phi}\tilde{E}_2/\tilde{E}_1+\delta_a}{\delta_a\delta_b+|f|^2}
\label{eq:1}
\end{eqnarray}

\begin{figure}[t!]
      \centering
        \includegraphics[height=2.1in]{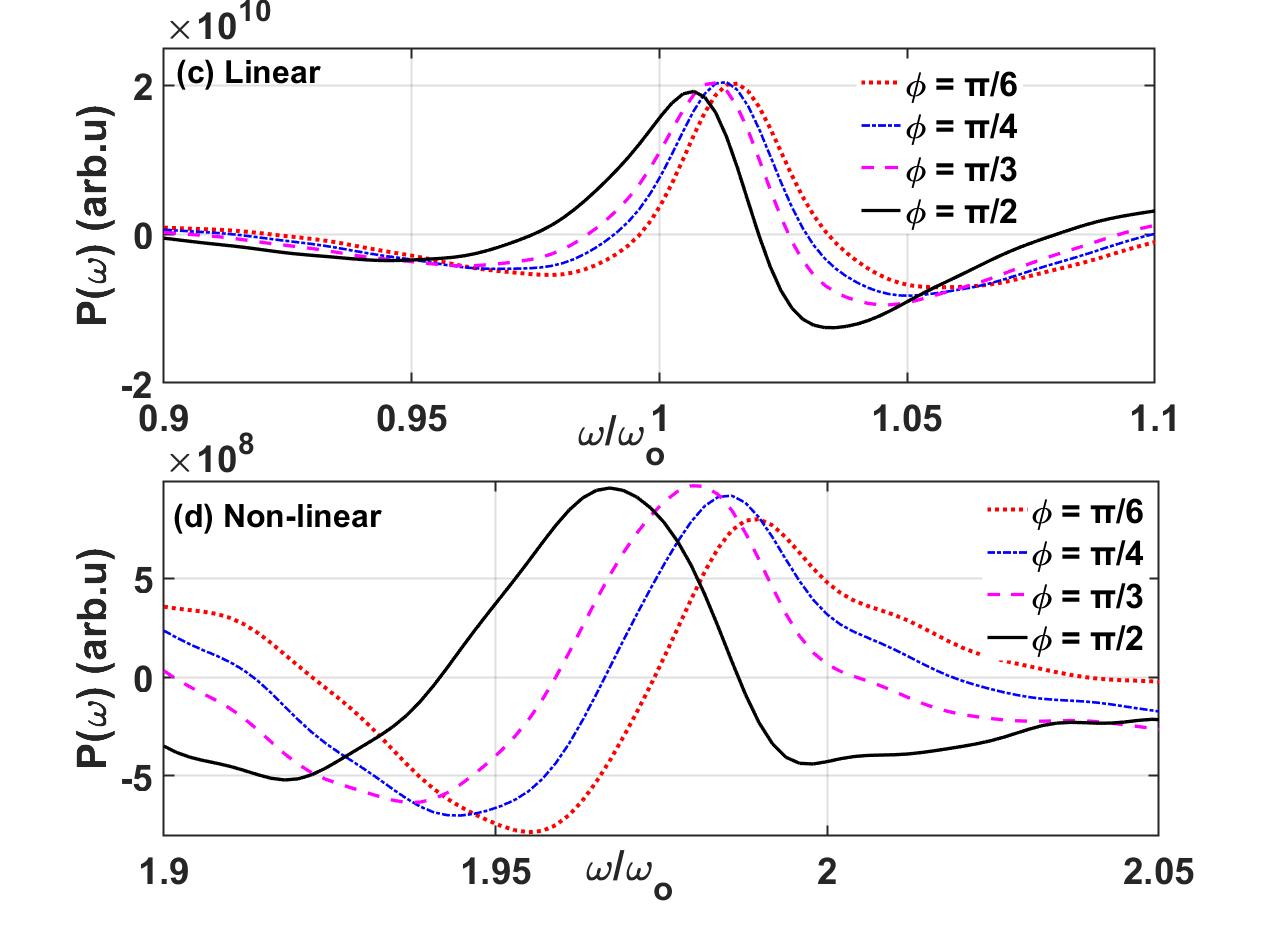}
    \caption{\label{fig:3} (a,b) Electric polarization in NE$_2$ demonstrating the intensity of linear and nonlinear plasmon mode as a function of normalized frequency for various phases ($\phi$) of control pulse. Results are evaluated through FDTD simulations.}
\end{figure}
where $\delta_a= [(\omega_a-\omega_o)+\gamma_a]$ and $\delta_b= [(\omega_b-\omega_o)+\gamma_b]$ consist of the resonances $(\omega_a=\omega_b=\omega_o)$ and corresponding decay rates $(\gamma_1=\gamma_2=0.05\omega_o)$ of nano-ellipsoids with $f$ as interaction strength between plasmon modes. The x-polarized pump pulse amplitude is in the real plane ($\tilde{E_2}$ $= E_2$) and y-polarized probe pulse amplitude in the imaginary plane ($\tilde{E_1}$ $= i E_1$). This choice does not affect the overall results or the underlying physics of the system, as the absolute values of plasmon amplitudes are used in the analysis. From Eq.\ref{eq:1}, it is seen that the susceptibility of the system depends on the phase difference $\phi$ between the pump and probe pulses. However, the tunability of probe field is solely governed by the enhancement of hotspot of $\alpha_i$ mode due to coupling with ($\beta_1$) mode at the intersection of NEs. The coupling strength is taken as $f=0.06\omega_o$ \cite{Panahpour2019}. The analytical results are computed and plotted in Fig.\ref{fig:2}. Figure.\ref{fig:2}(a) and \ref{fig:2}(b), demonstrate the real and imaginary parts of dielectric constant as a function of normalized frequency ($\omega_o$) for linear $\alpha_1$ and nonlinear $\alpha_2$ mode of NE$_1$. The curves show signature of plasmon analogue of index enhancement as suggested in \cite{Panahpour2019}. \textbf{At point A the imaginary part becomes zero while the susceptibility becomes negative at A$^{'}$, indicating zero or loss-less dispersion region. }This particular point indicates ENZ frequency region. In the linear case, ENZ frequency appear at \textbf{$\omega= 0.9272\omega_o$.}\\ 

\begin{figure}[t!]
        \centering
        \includegraphics[height=3in]{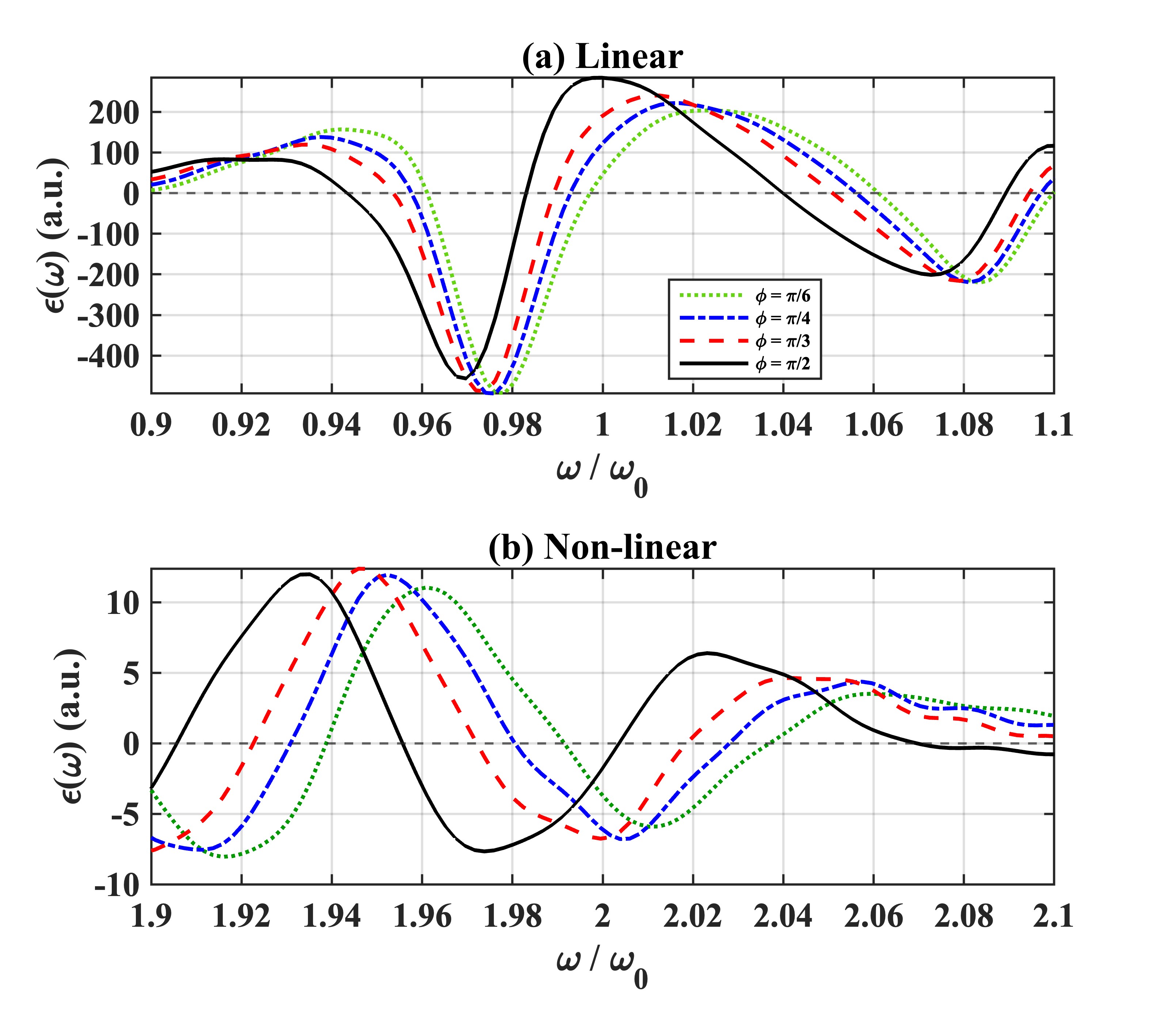}
    \caption{\label{fig:4} Real part of susceptibility $\chi(\omega)$ in the (a) linear and (b) nonlinear regime as a function of normalized frequency for various phases ($\phi$) of control pulse. Results are obtained using FDTD simulations. \textbf{(add imag parts here)}}
\end{figure}

For nonlinear system, we theoretically derive the nonlinear dielectric constant by solving the time-dependent complex amplitudes of linear and nonlinear plasmon mode (see theoretical calculations in SS). The frequency and decay rate parameters are set as $\Omega_a=\Omega_b=2\omega_o$ and $\Gamma=\gamma=0.05\omega_o$ with $f = 0.06\omega_o$ and $\chi^{(2)}=10^{-10}\omega_o$. The pump pulse amplitude is taken as E$_2=10$E$_1$. We calculate the dielectric constant for probe field derived from polarization induce in NE$_2$ for nonlinear system and plot the results in panel (b) of Fig.\ref{fig:2}. In the nonlinear system, the ENZ frequency is appear at \textbf{$\omega= 1.98\omega_o$} corresponding to the zero absorption. We aim to tune the ENZ frequency in both linear and nonlinear regime by changing the phase of pump pulse. At first, we investigate the shift in the spectral position of polarization for different phases of the control pulse. We perform both theoretical and FDTD calculations but here we only present FDTD simulation results. In FDTD setup, we simulate two Ag L-shaped NEs with similar configuration as shown in Fig.\ref{fig:1}. The geometrical parameters for NEs are taken from \cite{Panahpour2019} and the value of dielectric permittivity for silver NEs is obtained from Johnson and Christy \cite{Johnson1972}. To evaluate the x-and y-component of electric field polarization near the hotspot region, we placed time monitors between NEs. To analyze the impact of pump phase on the ENZ frequency, we evaluate the average electric polarization P$(\omega)$ and susceptibility $\chi(\omega)$ in NE$_2$ for comparison and plot the results, as shown in Fig.\ref{fig:3} (a,b)  and Fig.\ref{fig:4}(c,d), respectively.

In Fig.\ref{fig:3}, panels (a) and (b) show the spectral modulation in the polarization due to shift in energy at the hotspot for a specific phase of pump, while the amplitude of pump and probe pulse kept fixed at E$_2 =$ 10E$_1$. Tuning the control phase from $\pi/6$ to $\pi/2$ induces a significant shift in the linear resonances from \textbf{$1.15\omega_o$ to 1.0$\omega_o$} yielding \textbf{1.6$\%$} of modulation depth. Similarly, for nonlinear system the resonances blueshift from \textbf{$1.99\omega_o$ to $1.96\omega_o$} indicating coherent control of probe pulse. These spectral shifts are attributed to the indirect effect of changes at the hotspot of the probe, even though the pump source does not directly contribute to the polarization at the ends of the long axis of the vertical ellipsoid. Next, we inspect the shift in ENZ frequency for different phases by evaluating the electric susceptibility both in linear and nonlinear plasmonic system. 

Panels (a) and (b) in Fig.\ref{fig:4}, illustrate the real part of susceptibility for the linear ($\alpha_1$) and nonlinear ($\alpha_2$) modes of NE$_2$ and modulation in the ENZ resonances for different phases of control field (E$_2$). The color dots indicate the points of ENZ frequency when the real part of $\chi$ is zero and imaginary part of $\chi(\omega)$ is negative. At zero absorption region, the probe field propagation is minimum which enhances the localization and hence the hotspot intensity. Additionally at this particular frequency the \textbf{material permittivity becomes negative} which shows the emergence of ENZ region in nano-antenna structure. By varying pump phase from $\pi/6$ to $\pi/2$, we observe a prominent shift in the ENZ frequency corresponding t\textbf{o zero absorption point.} This shift also appear in the imaginary part of $\epsilon(\omega)$ indicating the control of probe's resonances at zero-index or negative extinction. In this way, we achieve a precise control over the tuned ENZ frequencies which stay fixed even if pump pulse is turned off. Such optical control of ENZ resonances with stable tunability holds promising potential to implement these structures in photonic systems. 
\begin{figure}
\includegraphics[scale=0.08]{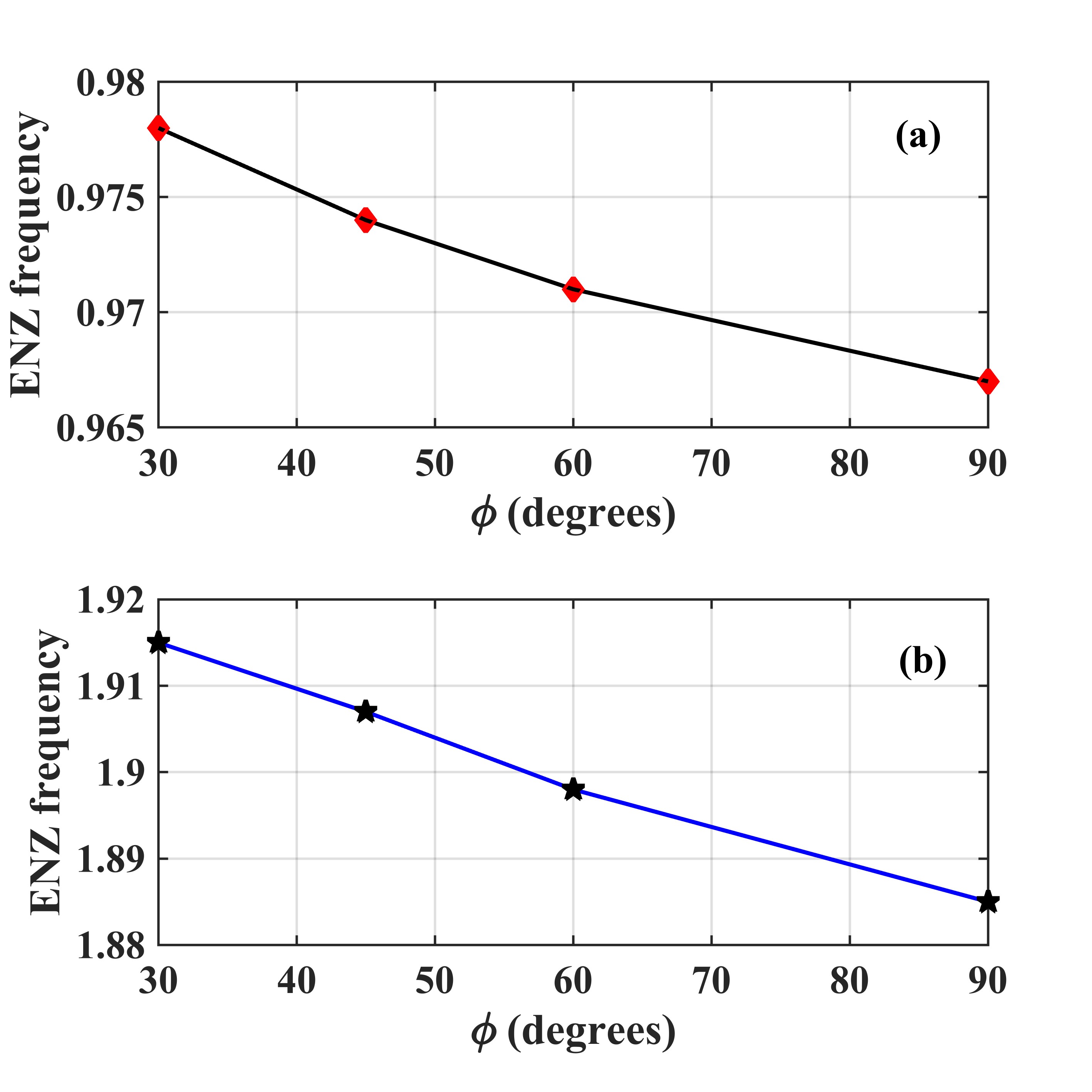}
\caption{\label{fig:5} Modulation in ENZ frequency in (a) linear and (b) nonlinear plasmonic systems for different control phases.}
\end{figure} 
To clearly indicate this precise shift, we plot the position of each ENZ frequency as a function of control phase for both linear and nonlinear case, as shown in Fig.\ref{fig:5}. Panels (a) and (b) demonstrate the active control of ENZ frequency at optimum phases. The significant spectral shift with control phases indicate that the optical tuning of operational bandwidth of the system is stable and reversible for broad spectrum of the control pulse. Also this could be beneficial for systems requiring optical modulation with desired ENZ wavelengths.\\
In summary, we proposed active tuning of ENZ resonances through phase modulation of pump pulse. Variation in the phase from $\phi=\pi/2$ to $\phi=\pi/6$ not only tunes the permittivity of probe but also induces a redshift in the system operational frequency in the visible regime. This resonant frequency's shift is also quite visible in \textbf{zero absorbance region, }which indicates that our proposed system can be employed as an active tunable near-zero-indexed or ENZ medium both in the linear and nonlinear quantum systems. Moreover, the coupling of two hotspot fields supported by L-shaped nanoantenna reduces the inherent plasmonic losses in metals, which is one of its more valuable aspect.\\
Our model demonstrates a proof of principle concept for on-demand coherent control of optical systems operating at visible frequencies for ENZ. Beyond providing a controllable mechanism for permittivity modulation, the ability to dynamically tune the ENZ resonant frequency offers new opportunities for applications in devices such an photonic integrated circuits (PICs), metasurfaces and quantum computing systems. Furthermore, active tuning of ENZ resonances and significant spectral tunability observed in our model system provides a precise control over light-matter interactions, enabling dynamical modulation of optical properties critical for developing reconfigurable photonic devices and next-generation optical systems.

\begin{acknowledgments}
E.O, H.A., M.G., M.E.T. and R.S. acknowledge support from TUBITAK No. 123F156.
\end{acknowledgments}
\bibliography{ENZ}

\end{document}